\RequirePackage{fix-cm}
\documentclass[smallextended]{svjour3}       
\smartqed  
\usepackage{graphicx}
\usepackage{color}
\usepackage{ulem}
\begin{document}

\title{Investigation of Supercurrent in the Quantum Hall Regime in Graphene Josephson Junctions\thanks{A. Draelos and M. T. Wei contributed equally to this work.}
}

\titlerunning{Supercurrent in the Quantum Hall Regime in Graphene Josephson Junctions}        

\author{Anne W. Draelos \and
	Ming Tso Wei \and
	Andrew Seredinski \and
	Chung Ting Ke \and
	Yash Mehta \and
	Russell Chamberlain \and 
	Kenji Watanabe \and
	Takashi Taniguchi \and
	Michihisa Yamamoto \and
	Seigo Tarucha \and
	Ivan V. Borzenets \and
        Fran\c cois Amet \and
        Gleb Finkelstein
}

\institute{A. W. Draelos \and M. T. Wei \and A. Seredinski \and C. T. Ke \and G. Finkelstein \at
              Department of Physics, Duke University, Durham, North Carolina 27708, USA \\
               \email{amw73@duke.edu, ming.tso.wei@duke.edu}
		\and
	     K. Watanabe \and T. Taniguchi \at
	     Advanced Materials Laboratory, NIMS, Tsukuba 305-0044, Japan
	     \and
	     M. Yamamoto \at
	     PRESTO, JST, Kawaguchi-shi, Saitama 332-0012, Japan
	     \and
	     S. Tarucha \at
	     Department of Applied Physics, University of Tokyo, Bunkyo-ku, Tokyo 113-8656, Japan\\
	     Center for Emergent Matter Science (CEMS), RIKEN, Wako-shi, Saitama 351-0198, Japan
	     \and
	     I. V. Borzenets \at
	     Department of Physics, City University of Hong Kong, Kowloon, Hong Kong SAR
	     \and
	     Y. Mehta \and R. Chamberlain \and F. Amet \at
	     Department of Physics and Astronomy, Appalachian State University, Boone, North Carolina 28607, USA
}

\date{Received: date / Accepted: date}

\maketitle

\begin{abstract}
In this study, we examine multiple encapsulated graphene Josephson junctions to determine which mechanisms may be responsible for the supercurrent observed in the quantum Hall (QH) regime. Rectangular junctions with various widths and lengths were studied to identify which parameters affect the occurrence of QH supercurrent. We also studied additional samples where the graphene region is extended beyond the contacts on one side, making that edge of the mesa significantly longer than the opposite edge. This is done in order to distinguish two potential mechanisms: a) supercurrents independently flowing along both non-contacted edges of graphene mesa, and b) opposite sides of the mesa being coupled by hybrid electron-hole modes flowing along the superconductor/graphene boundary. The supercurrent appears suppressed in extended junctions, suggesting the latter mechanism.
\end{abstract}

\section{Introduction}
\label{intro}

\begin{figure*}
  \includegraphics[width=\textwidth]{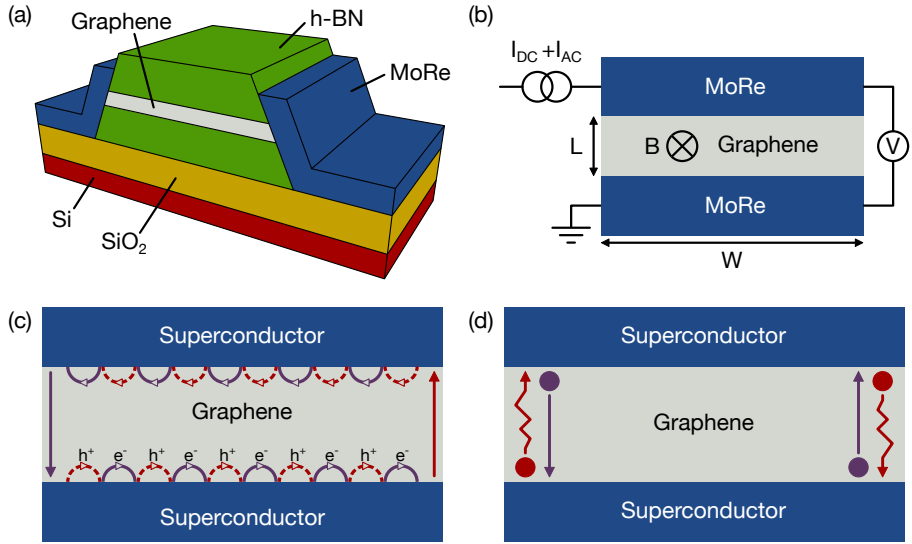}
\caption{(a) A schematic of the graphene Josephson junction stack. The graphene is sandwiched in between hexagonal boron nitride (h-BN) flakes and contacted along the edge by molybdenum-rhenium (MoRe) leads. The entire ensemble is placed atop a conductive Si substrate with a 300 nm oxide layer to serve as the back gate. (b) Schematics of the four-terminal measurement setup. Both AC and DC current bias are applied to one end of the junction while the other end is grounded. Voltage probes then measure the voltage difference across the junction. (c) A diagram showing the Andreev bound states encompassing the entire graphene perimeter. The electron and hole chiral modes on the opposite sides of the sample are connected by the theoretically predicted hybrid electron-hole modes propagating along the superconducting interface. (d) A diagram picturing an alternative mechanism, where each  vacuum edge of the graphene mesa supports supercurrent independently, due to spurious channels potentially existing along the edges.}
\label{fig:1}       
\end{figure*}

The recent development of type-II superconducting contacts to heterostructures of graphene encapsulated in boron nitride has enabled the fabrication of microns-long, ballistic devices, which host a variety of novel electronic phenomena \cite{Calado,Shalom,Allen,Lee,Amet,Park}. The junctions in this study are made of two superconducting electrodes separated by a region of encapsulated graphene. The transmission of charge carriers across the normal-superconducting interface relies on the Andreev reflection process, where an incoming electron is retroreflected as a hole at the interface while a Cooper pair is transmitted to the superconductor \cite{Nazarov}. A normal channel between two superconducting electrodes hosts Andreev bound states (ABS) that arise from coherent Andreev reflections at both interfaces.  At high magnetic fields, graphene enters the quantum Hall regime and the gapped bulk prohibits regular supercurrent flow through the junction. The Landau level quantization in graphene is exhibited by plateaus of resistance as a function of density that correspond to constant filling factors ($v=nh/eB=\pm2,6,10, \ldots$). Recently, our experimental evidence has shown that supercurrent could still flow in this regime \cite{Amet}.

In similar structures formed from topological insulators, helical states flow along each edge in both directions so that both edges can support ABSs and carry supercurrent independently \cite{Hart}. In contrast, because of the cyclotron motion, each edge of the QH junction conducts charge carriers (both electrons and holes) in one direction. Nevertheless, theory predicted that supercurrent could still be conducted in the QH regime by the counter-propagating edge states formed on the opposite sides of the sample \cite{Ma,van Ostaay,Stone}. In Ref. \cite{van Ostaay}, it has been proposed that these edges are connected by the hybrid electron/hole mode propagating along the superconductor-graphene interfaces (Fig. 1c) \cite{Hoppe}, forming Andreev bound states that encompass the entire junction perimeter. Namely, an electron could travel across the length of the junction along one edge, then couple to a hybrid mode which traverses the contact to the opposite edge, and then couple to the hole traveling back along that edge \cite{van Ostaay}. However, one may alternatively hypothesize that electrons or holes could propagate along the etched edges of the mesa against the chiral direction, as it happens at lower fields (see Supplementary Information in Ref \cite{Shalom}), leading to a more traditional form of ABSs on each edge (Fig. 1d).

In all of our junctions, supercurrent in the QH regime showed strong periodic modulations versus magnetic field (see for example, Fig. 3). The measured period was close to the one observed at small field, $h/2e$ (Fraunhofer-like pattern; see Fig. 2c). For an ABS encompassing the full sample circumference, with a hole propagating along one edge and an electron along the other, the period has been predicted to be $h/e$, twice that observed at small fields \cite{van Ostaay}. This discrepancy raises the question whether instead supercurrent propagates along each edge separately, resulting in a SQUID-like behavior with a periodicity of $h/2e$.

To further elucidate the mechanism by which supercurrent flows in the QH regime, we examine the influence of the junction shape and dimensions on the strength of the supercurrent and its location with respect to the gate voltage (i.e. filling factor). In the first part of the paper (Sec. 3.1-2), we compare several junctions of constant widths and different lengths, or vice versa, made of the same graphene crystal. In the second part (Sec. 3.3), we present results measured on junctions in which one of the edges of the mesa is extended beyond the ends of the contacts, while the other edge is kept short. If, as posited in an earlier work \cite{Amet}, the Andreev bound states enclose the entire perimeter of the junction (see Fig. 1c), the supercurrent will be suppressed in the extended junctions compared to a standard rectangular devices. In contrast, if ABS are formed along each edge of the junction due to trivial edge states, the short edge of the junction will still support supercurrent in extended samples. 
 
\section{Experimental Details}
\label{sec:1}
	All devices are fabricated in the same manner. Graphene monolayer flakes are exfoliated from Kish graphite and encapsulated with hexagonal boron nitride (h-BN) flakes using a standard stamping method \cite{Wang}. These stacks are then deposited onto p-doped silicon wafers that have a 300 nm thermally grown oxide layer on top. A thermal anneal at 500 $^\circ$C follows, producing a clean, bubble-free section of the graphene mesa that can be used for defining the junctions. Raman scattering and Atomic Force Microscopy measurements are performed to confirm the quality and single-layer nature of the encapsulated graphene. The stack is patterned using standard electron-beam lithography and then reactive-ion etched using a CHF$_{3}$/O$_{2}$ plasma to expose the edges of the graphene. Molybdenum-rhenium (MoRe) alloyed contacts are deposited using DC magnetron sputtering with a thickness of about 100 nm. MoRe is a type II superconductor with a critical field of 8 T and $T_c$ of about 8 K \cite{Calado}. It makes excellent electrical contact with graphene in a Josephson junction, with a measured gap of roughly 1.2 meV \cite{Amet}.
		
\begin{figure*}
  \includegraphics[width=\textwidth]{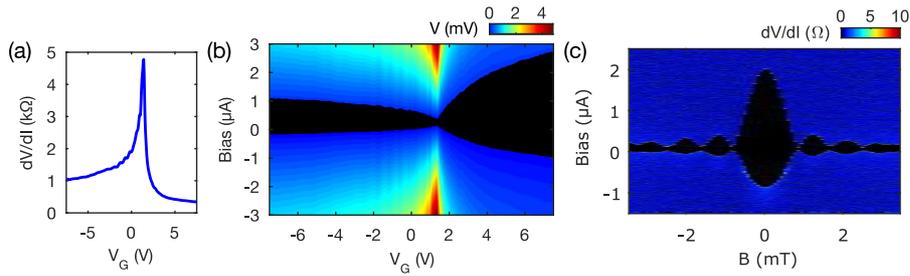}
\caption{Characteristics of ballistic transport in a graphene Josephson junction (device C5 J3). (a) Differential resistance as a function of back gate voltage showing sharp, narrow Dirac peak. Oscillations in the resistance on the p-doped side result from partial reflections from the PN interface formed close to the superconducting contacts that n-dope the adjacent graphene. (b) Voltage across the junction as a function of back gate voltage taken at base temperature ~30 mK. The superconducting region of zero resistance (black) is seen at zero DC current bias and up to the critical current when it switches to the normal resistive state. (c) Conventional Fraunhofer-like interference pattern measured under small magnetic fields at high electron-doping showing a uniform current distribution in the junction.}
\label{fig:2}       
\end{figure*}

The samples are cooled in a top-loading dilution refrigerator (Leiden Cryogenics) with a previously confirmed electron base temperature of $\sim$ 30 mK (in the normal state). The devices are isolated via resistive coaxial lines, low temperature two-stage RC and metal powder filters, as well as RF shielding \cite{Ke}. Transport measurements were conducted in a standard four-terminal setup (Fig. 1b) with a back gate voltage V$_{G}$ controlling the carrier density in the graphene. Each junction has two MoRe contacts on either side that diverge into two leads past the graphene. Typically, a small AC excitation current of $\sim$ 50 pA is used to measure the differential resistance of the junction around zero bias; a large DC bias is added to switch the junction from a superconducting to a normal state. Magnetic field is applied perpendicular to the sample plane.

The Josephson junctions' quality and ballistic properties are evaluated through a series of standard measurements. As the back gate voltage is swept through the Dirac point of the graphene, a narrow and precipitous spike in the four-probe resistance of the junction should appear as a principle transport signature (Fig 2a). Ballistic transport implies a sample resistance independent of junction length, so a set of graphene junctions of varying lengths demonstrates a resistance per contact width independent of junction length. No trend towards higher resistance is observed in junctions up to 2 $\mu$m in length \cite{Borzenets}.

The normal state conductance of the samples at high electron density is comparable to the ballistic limit, indicating high contact transparency with the full transmission coefficient across the junction estimated to exceed 0.9 \cite{Amet,Borzenets}. Note that the resistance for electron doping is lower than that for hole doping. Graphene directly adjacent to MoRe is electron-doped due to the work function mismatch between the two. When the junction is gated to the hole-doped regime, two p-n interfaces are formed, which result in a nonzero reflection probability for holes in the junction. In a ballistic sample, phase-coherent transport through this cavity leads to a Fabry-Perot (FP) interference pattern as the Fermi wavevector is tuned through a series of resonances \cite{Shalom,Miao,Young,Rickhaus}. These oscillations can be seen in the resistance as well as the switching current of the junction visible on the hole-doped side of a bias-gate map (Figure 2a,b).

\begin{figure*}
  \includegraphics[width=\textwidth]{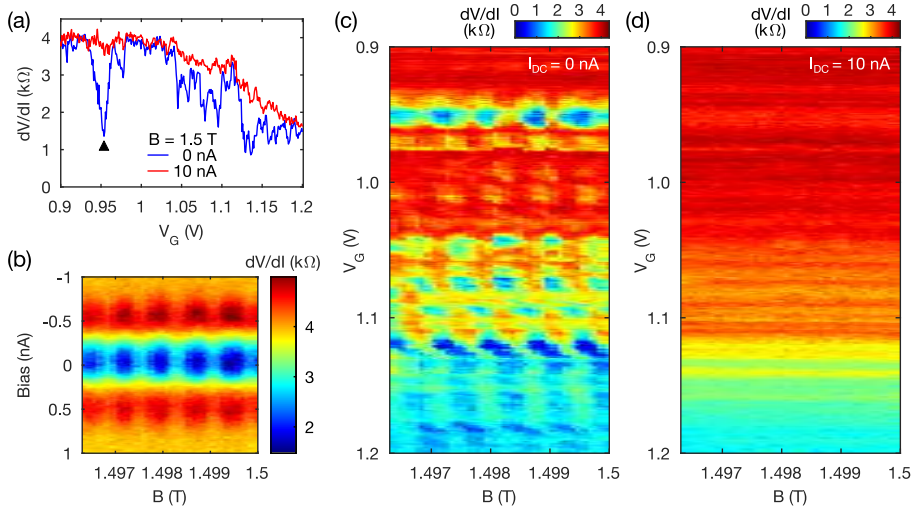}
\caption{Magnetic interference patterns of supercurrent in the QH regime on the device C5 J3. (a) Differential resistance $dV/dI$ measured at zero bias (blue) and 10 nA (red)  at the $\nu$=6 plateau. (b) Bias-field map showing periodic supercurrent at $V_G$=0.95 V, marked in (a) with a black triangle. (c)-(d) Maps of $R$ measured at zero and 10 nA current bias, respectively, as a function of gate voltage and field. All superconducting pockets in (c) show the same magnetic field periodicity. The periodic patterns are suppressed in the high bias case.}
\label{fig:3}       
\end{figure*}

The supercurrent in a high quality junction with a contact width on the order of microns easily reaches the range of a few $\mu$A. This supercurrent is measured by sweeping a DC current from negative to positive bias, showing the hysteresis between the retrapping and switching transitions into and out of the superconducting state. The uniformity of its distribution across the junction width, another indication of device cleanliness, can be confirmed by measuring its dependence on magnetic field. Bias sweeps through the superconducting branch of the I-V curve are run while a small magnetic field up to a few mT is applied perpendicular to the junction. The resulting Fraunhofer interference pattern, like that shown in Figure 2c, can be Fourier-transformed to show the current distribution of the junction \cite{Allen,Hart,Dynes}. In this case, the regular Fraunhofer-like pattern indicates a uniform current distribution across the width of the contacts, with a periodicity as expected from the device geometry, taking into account an effective flux focusing area due to field expulsion from the superconducting contacts \cite{Rosenthal}. 
                
In the quantum Hall regime, supercurrent with an apparent magnitude of $\sim 1$ nA is observed \cite{Amet}. This greatly reduced value compared to zero field could be explained by the following factors: 1) While at zero field the current flows across the full width of the contacts, at high field the current is carried only by a few edge channels. 2) The length of the Andreev bound states is at least equal to twice the length of the junction (Fig. 1d), or equal to the full perimeter of the junction (Fig. 1c), placing the junctions in the long junction regime \cite{Kulik,Bardeen,Svidzinsky,Svidzinsky2,Golubov}. These factors are expected to bring the supercurrent to the nA range. The critical current in the few to tens of nA range is notoriously difficult to measure \cite{Steinbach,Joyez,Vion,Jarillo}. It is quite likely that environmental and thermal fluctuations reduce it to the apparent switching current on the 1 nA scale, as observed here. Thermal activation measurements conducted in the Supplementary Materials of Ref. \cite{Amet} indicate that the true critical current in this regime in typically is in the range of a few nA.

Figure 3 presents several ways of visualizing this supercurrent by measuring the differential resistance of the junction, $R=dV/dI$. First, the pockets of supercurrent appear as reproducible dips in $R(V_G)$ curves measured at zero bias (Fig. 3a). These dips are not seen in the corresponding sweeps measured at bias which is high enough to overcome supercurrent ($\sim$ a few nA). Once supercurrent on a QH plateau is tentatively identified, its full $R(I)$ curve is measured. Figure 3b shows $R(I)$ measured as a function of magnetic field for a prominent spot at $V_G=0.95$ V in Figure 3a. The supercurrent is clearly periodic in magnetic field, and its periodicity is very close to the one observed in the Fraunhofer pattern (Fig. 2c). This period is constant across a wide range of magnetic fields and filling factors. Finally, the map of $R(V_G,B)$ allows one to measure the periodicity over a wide range of gate voltages (Fig. 3c). The periodic features associated with the supercurrent are suppressed by application of the bias current of $10$ nA (Fig. 3d).
                
\section{Results and Discussion}
\begin{table}[h]
\caption{The eight junctions from four different samples and their dimensions.}
\label{tab:1}       
\begin{tabular}{lll}
\hline\noalign{\smallskip}
Junction & Width ($\mu m$) & Length (nm)  \\
\noalign{\smallskip}\hline\noalign{\smallskip}
C5 J2 & 3 & 1000 \\
C5 J3 & 3 & 400 \\
C5 J4 & 3 & 200 \\
F2 J3 & 4.5 & 500 \\
F2 J4 & 1.5 & 500 \\
E2 J2 & 2 & 500 \\
E2 J4 (extended) & 2 + & 500 \\
G1 J1 (extended) & 3 + & 500 \\
\noalign{\smallskip}\hline
\end{tabular}
\end{table}
In total, eight junctions on four different graphene crystals were studied. Sample C5 was composed of rectangular junctions of different lengths and same width; sample F2 had 2 rectangular junctions of different widths and same length, as summarized in Table 1. Samples E2 and G1 had one graphene edge extended past the contact interfaces by a few microns (see Fig. 6a). All junctions (both rectangular and extended edge) showed clean ballistic behavior with microamperes of supercurrent at base temperature and in zero field. 

\subsection{Length Dependence}
\label{sec:2}

\begin{figure*}
  \includegraphics[width=\textwidth]{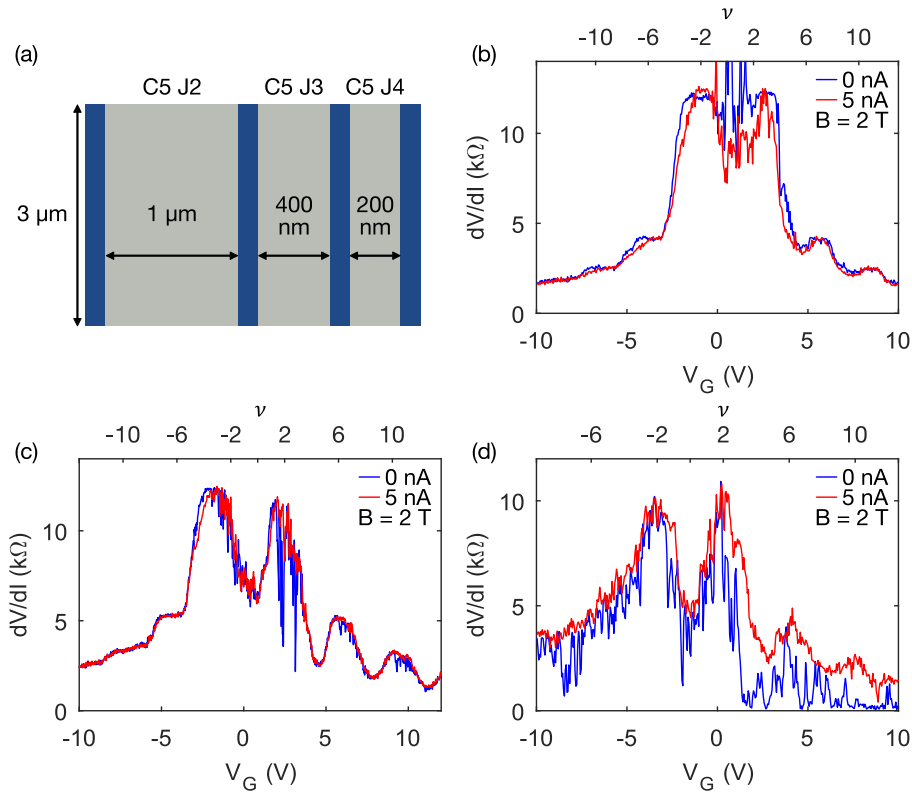}
\caption{Length dependence of supercurrent at 2 T. (a) Schematic of the three junctions on one graphene mesa, not drawn to scale. (b)-(d) Differential resistance as a function of back gate voltage measured with zero (blue) or high (red) applied DC current bias to suppress supercurrent for C5 J2, C5 J3, and C5 J4, respectively. Dips in the resistance show pockets of superconductivity living on or between the QH plateaus.}
\label{fig:4}       
\end{figure*}

We measured three junctions (C5 J2, C5 J3, C5 J4) on the same graphene crystal that had the same width but different lengths (200-1000 nm). These junctions have been studied in zero magnetic field as junctions A-C in Ref. \cite{Borzenets}.  Figure 4 shows differential resistance vs. $V_G$ measured at 2 Tesla with an AC current of 50 pA. A DC current bias of zero (blue) and 5 nA (red) was applied in addition to the AC excitation. 5 nA is sufficient to suppress supercurrent through graphene in high fields, and this measurement is expected to reveal mostly the QH features. On the other hand, at zero DC bias we see a suppressed resistance due to superconductivity. Note that the measured I-V curves in the supperconducting regime show a non-zero differential resistance; this is attributed to phase diffusion as the estimated Josephson energy is high but comparable to the temperature \cite{Amet,Kautz,Ingold,Borzenets2}.

The behavior of the junctions in the QH regime shows a simple trend. At 2 T, the cyclotron diameter is shorter than the length of all junctions, so they are beyond the semiclassical limit. Plateaus are visible under high bias conditions when the supercurrent through graphene is suppressed, though their shapes of $R$ vs. $V_G$ curves vary due to the different aspect ratios of the junctions \cite{Williams}. The two shorter junctions (C5 J4 and C5 J3) display supercurrent (Fig 4c,d  where the blue curve is suppressed compared to the red curve). In contrast, the longest junction (C5 J2) has no visible supercurrent. Moreover, the shortest junction (C5 J4) exhibits many more superconducting pockets with greater depth compared to the next junction (C5 J3).  However, this behavior is consistent with both possible mechanisms of the supercurrent: the one requiring circumferential ABS and the one relying on the existence of trivial states along the etched edges of the mesa.

\subsection{Width Dependence}
\label{sec:3}

\begin{figure*}[b]
  \includegraphics[width=\textwidth]{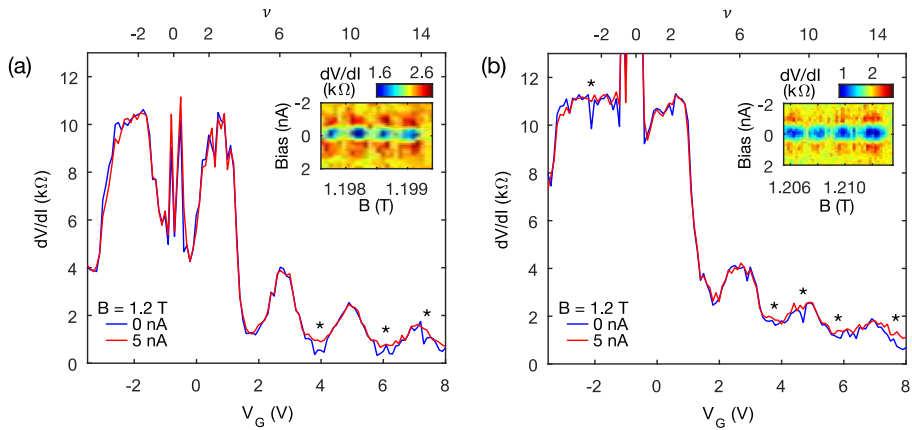}
\caption{Width dependence of supercurrent in the quantum Hall regime. (a)-(b) Differential resistance as a function of back gate voltage measured with 0 (blue) or 5 (red) nA applied DC current bias on F2 J3 and F2 J4, respectively. Dips in the resistance show pockets of superconductivity living on or between the QH plateaus. (Insets) Magnetic field interference patterns at gate locations where a superconducting pocket exists for F2 J3 and F2 J4, respectively. Periodicity is remarkably the same for each junction as it was at near-zero fields.}
\label{fig:5}       
\end{figure*}

We next compare two adjacent junctions (F2 J3 and F2 J4) made on the same graphene mesa with the same junction length but with widths differing by a factor of three (see Table 1). The overall graphene perimeters are different by a factor of 2.5 between the two. As the graphene enters the quantum Hall regime, pockets of superconductivity again appear at certain gate voltages. Figure 5 shows the differential resistance of the junctions under zero and 5 nA DC current bias as a function of the back gate voltage V$_{G}$. Measurements were performed at 1.2 T or higher, sufficiently deep in the QH regime.

Superconducting pockets can be seen in both wide and narrow junctions (marked with black stars in Fig. 5). They are located both on top of plateaus (v=2,6) as well as in the dips between the plateaus. The periodicity seen in either junction (0.6 mT and 1.8 mT for F2 J3 and F2 J4, respectively (Fig. 5 insets)) is consistent with the above mentioned devices at zero field. Interestingly, when comparing the number and depth of superconducting pockets between the wide and narrow junctions, we find two seemingly opposing trends. At 1.4 T, the narrow junction has a greater number of pockets, with more of them located in the middle of the quantized plateaus. Furthermore, the junction exhibits supercurrent at lower filling factors ($v=2,6$) which correspond to higher resistances, ruling out a simple relationship between low resistance and the appearance of supercurrent. However at slightly higher fields, around 1.7 T, all of the supercurrent in the narrow junction has died out while the $B$-periodic supercurrent in the wide junction is still visible. These characteristics of the supercurrent do not seem to show an understandable dependence on the width of the contacts.

\subsection{Samples with Extended Edges}
\label{sec:4}

\begin{figure*}[b]
  \includegraphics[width=\textwidth]{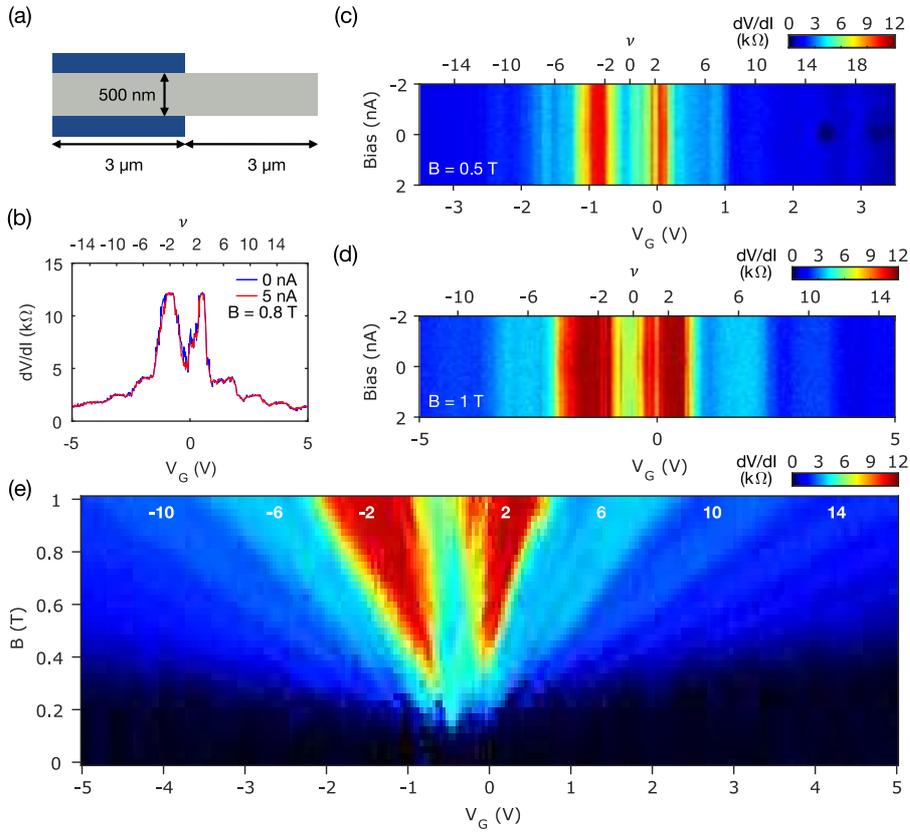}
\caption{(a) Diagram of an extended edge sample (G1 J1), not to scale. (b) Differential resistance as a function of back gate voltage measured with 0 (blue) or 5 (red) nA applied DC current bias at 0.8 T. Dips in the resistance still show remnants of superconductivity. (c-d) Bias-gate maps showing supercurrent at 0.5 T and no supercurrent at 1 T, respectively. (e) Fan diagram of the differential resistance measured at zero applied bias versus gate voltage and magnetic field from 0 to 1 T. Supercurrent at small fields is not suppressed, but in the QH regime the superconducting pockets appear only up to about 0.8 T.}
\label{fig:6}       
\end{figure*}

To further investigate the origin of supercurrent in the QH regime, we next studied two devices with similar dimensions to the rectangular samples presented above, but with one edge of the graphene mesa extended past the superconducting contacts as pictured in Figure 6a. The extended region lengthens one of the graphene edges by 2 to 4 microns in these two samples. The junctions have the superconducting contacts separated by 500 nm, well within the range of distance over which supercurrent was observed in the rectangular samples. Both extended junctions have one edge of the mesa defined in exactly the same way as the edges of the rectangular junctions. If the supercurrent flows along some unidentified trivial edge states, the short edge of the mesa is expected to support supercurrent of comparable magnitude, $\sim 0.1-1$ nA, to the rectangular samples. The extended edge of the mesa is expected to be much too long to support supercurrent, so the signal would not be periodic in magnetic field. If, on the other hand, the supercurrent is explained by the putative perimeter-encircling ABS which involves both edges, we expect it to be suppressed in the extended samples.

The signatures of the supercurrent observed in high fields indeed differ greatly between standard and extended samples. Figure 6b shows the gate sweeps at zero and 5 nA bias measured at $B=0.8$ T on one of the extended junction (G1 J1), which has an additional 3 micron-long rectangular appendage off to one side. This junction has heavily suppressed supercurrent in high field, apparently only surviving at one pocket at $B=$0.8 T. While at $B=0.5$ T a bias-gate map shows several pockets of supercurrent residing on quantized plateaus (Fig. 6c), by 1 T all evidence of supercurrent vanishes (Fig 6d). An extensive study of the observed superconducting pockets in G1 J1 showed no periodic behavior. It should be noted that the quality of graphene in this junction is comparable to the junctions presented previously, and the supercurrent at zero and low fields is not suppressed (Fig. 6e). Comparatively, multiple rectangular junctions of comparable sizes ($W=2-3$ microns by $L=500$ nm) show pockets of superconductivity well into the quantum Hall regime ($\sim$2 T). This result indicates that any possible supercurrent carried by a single edge vanishes at $B< 1$ T. The supercurrent observed at higher fields in rectangular junctions should be then attributed to a different mechanism that requires both edges to be present simultaneously, such as the ABS encompassing the whole junction perimeter.

The second sample (E2 J4) has the mesa extended by several microns in an 'L' shape (schematics in Fig. 7). In this junction, the supercurrent is also suppressed by $B=1$ T in the gate voltage range corresponding to the QH regime, thus supporting the observations made in Figure 6. However, at a lower field of 0.7 T we observe supercurrent which unexpectedly is periodic in magnetic field (Fig. 7b-d). Notably, the observed supercurrent has a period of about 1.4 mT, which is close to the 1 mT periodicity observed in a comparable rectangular junction (E2 J2). The much greater area of E2 J4 compared to E2 J2, however, should have resulted in an interference pattern with a period of $\sim 0.3$ mT. The observed periodicity suggests that the supercurrent is formed in the region of the mesa between the contacts, excluding the extended region.

\begin{figure*}
  \includegraphics[width=\textwidth]{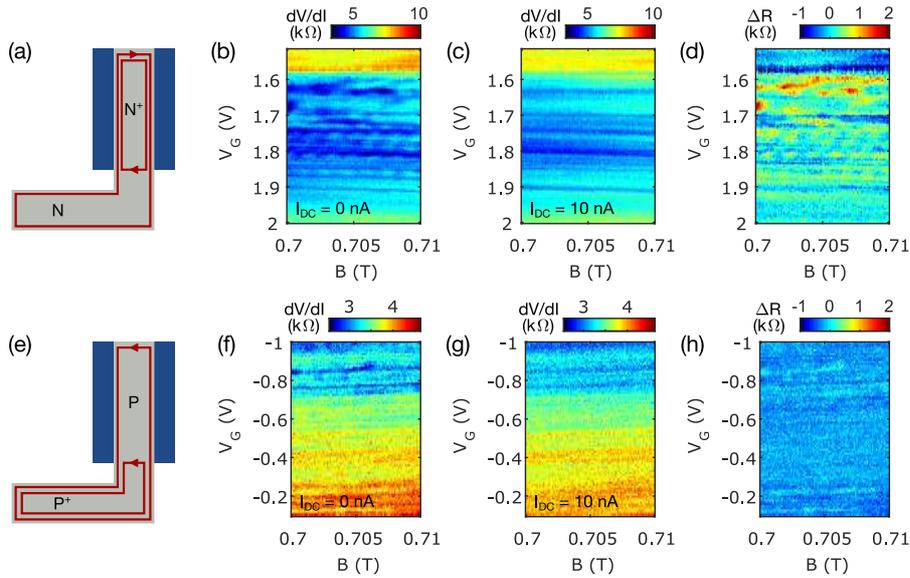}
\caption{Supercurrent in the 'L'-shaped extended edge sample at 0.7 T. (a),(e) Diagrams depict edge modes formed due to the density differences between different parts of the graphene mesa. (b-c) Back gate - field maps taken in the electron doping regime (including $\nu=2,6$) under zero (f) and 10 nA (g) of current bias. (d) The difference of the zero-bias and finite bias resistances, $\Delta R$ shows strong periodic modulations. (f-h) Back gate - field maps taken in the hole doping regime (including $\nu=-6,-10$) under zero (b) and 10 nA (c) of current bias. (h) Here, the difference of the zero-bias and finite bias resistances, $\Delta R$, shows no signs of periodicity. }
\label{fig:7}       
\end{figure*}

To understand this behavior, we note that based on the position of the Dirac point the bulk of this junction is P-doped, while the region next to the MoRe contacts is N-doped. When positive gate voltage is applied, this density difference will persist: the region close to the contacts will have a higher density of electrons (N$^+$) compared to the extended region (N). As a result, at some gate voltage range, edge states may form at the N-N$^+$ boundary (Fig. 7a). We surmise that these states participate in the ABS encircling the graphene region between the contacts, which enables the observed supercurrent with the period roughly corresponding to that area. Importantly, the states formed at N-N$^+$ boundary and the states formed at the short edge of the mesa on the other side of the contacts should flow in opposite directions, allowing the formation of the ABS. 

On the other hand, at negative gate voltages, the region close to the contacts will have a lower density of holes (P) compared to the extended region (P$^+$). As a result, the edge states at the P-P$^+$ boundary should flow in the same direction as the edge state on the opposite side of the contacts, and the ABS could not form (Fig. 7e). Indeed, we do not observe any signs of the periodic supercurrent at the comparable hole doping (Fig. 7f-h). At the same time a reference rectangular junction (E2 J2) shows strong periodic modulations of supercurrent at the comparable hole density and the same field.

\section{Conclusion}
\label{sec:5}
In this paper we have discussed the varied signatures of supercurrent transport in the QH regime. Rectangular junctions of various lengths and widths all show supercurrent in high fields, and the interference pattern in high magnetic field exhibits periodicity very close to the Fraunhofer pattern measured near zero field.  Supercurrent is efficiently suppressed by increased distance between contacts, and no supercurrent in the QH regime have been observed for a distance of $\sim2$ microns. The typical magnitude of supercurrent seems to show little dependence on the width of the contacts. Simplistically, these trends may suggest that supercurrents are flowing independently on each edge of the sample in a SQUID-like fashion. However, the studies of two junctions with extended mesa suggest that at least in those junctions a single short edge is not capable of supporting supercurrent beyond few hundred mT, lending credence to the alternative mechanism of supercurrent originating from the ABS encompassing the entire sample perimeter. Clearly, further studies are needed pinpoint the exact mechanism of the supercurrent in the quantum Hall regime.

\begin{acknowledgements} 
Low-temperature electronic measurements performed by A.W.D., M.T.W., C.T.K., and G.F. were supported by ARO Award W911NF-16-1-0122. Sample fabrication and characterization conducted by M.T.W., A.S., C.T.K., and G.F. were supported by the Division of Materials Sciences and Engineering, Office of Basic Energy Sciences, U.S. Department of Energy, under Award DE-SC0002765. F.A. acknowledges the ARO under Award W911NF-16-1-0132. I.V.B. and M.Y. acknowledge the Canon foundation. This work was performed in part at the Duke University Shared Materials Instrumentation Facility (SMIF), a member of the North Carolina Research Triangle Nanotechnology Network (RTNN), which is supported by the National Science Foundation (Grant ECCS-1542015) as part of the National Nanotechnology Coordinated Infrastructure (NNCI). M. Y. acknowledges support from JSPS KAKENHI Grant No. JP25107003. K. W. and T. T. acknowledge support from JSPS KAKENHI Grant No. JP15K21722 and the Elemental Strategy Initiative conducted by the MEXT, Japan. T. T. acknowledges support from JSPS Grant-in-Aid for Scientific Research A (No. 26248061) and JSPS Innovative Areas $\lq$Nano Informatics$\rq$ (No. 25106006). M. Y. and S. T. acknowledge support by Grant-in-Aid for Scientific Research S (No. 26220710), and Grant-in-Aid for Scientific Research A (No. 26247050, 16H02204).
\end{acknowledgements}



\end{document}